\documentclass[onecolumn]{revtex4}
\usepackage{graphicx}

\begin{document}

\title{CREDO project\footnote{Presented at Matter To The Deepest
Recent Developments In Physics Of Fundamental Interactions
XLIII International Conference of Theoretical Physics, September 2019, Katowice, Poland}
}

\author{Robert Kami\'nski$^a$, Tadeusz Wibig$^b$, 
David Alvarez Castillo$^a$, 
Kevin Almeida Cheminant$^a$, 
Aleksander \'C†wik\l{}'a$^c$, 
Alan R. Duffy$^d$, 
Dariusz G\'ora$^a$,
Piotr Homola$^a$, 
Pawe\l{}' Jagoda$^e$,
Marcin Kasztelan$^f$, 
Marek Knap$^g$,
Konrad Kopa\'nski$^a$, 
Peter Kovacs$^h$, 
Micha\l{} Krupi\'n"ski$^a$, 
Marek Magry\'s›$^i$, 
Vahabeddin Nazari$^j$, 
Micha\l{}' Nied\'zwiecki$^c$,
Wojciech Noga$^a$ , 
Mat\`ias Rosas$^k$, 
Szymon Ryszkowski$^g$, 
Katarzyna Smelcerz$^c$, 
Karel Smolek$^l$,
Jaros\l{}aw Stasielak$^a$,
S\l{}'awomir Stuglik$^a$, 
Mateusz Su\l{}'ek$^g$
Oleksandr Sushchov$^a$, 
Krzysztof Wo\'zniak$^a$}

\affiliation{
\vspace*{0.5cm}
$^a$Institute of Nuclear Physics PAS, Radzikowskiego 152, Krak\'ow,\\ 
$^b$University of Lodz, Faculty of Physics and Applied Informatics, 90-236 \L{}\'od\'z, Pomorska 149/153, Poland,\\
$^c$Cracow University of Technology, Krak\'ow, Poland \\
$^d$Swinburne University of Technology, Melbourne, Australia \\
$^e$AGH University of Science and Technology, Krak\'ow, Poland \\
$^f$National Centre for Nuclear Research, Otwock-\'Swierk, Poland \\
$^g$CREDO Collaboration, Poland \\
$^h$Wigner Research Centre for Physics, Budapest, Hungary \\
$^i$Academic Computer Center Cyfronet, AGH University of Science and
Technology, Krak\'ow, Poland \\
$^j$JINR Dubna, Russia \\
$^k$CREDO Collaboration, Uruguay \\
$^l$IEAP, Czech Technical University in Prague, Czech Republic}

\begin{abstract}
The Cosmic-Ray Extremely Distributed Observatory (CREDO) is a project created a few years ago in the Institute of Nuclear Physics PAS in Krak\'ow 
and dedicated is to global studies of extremely extended cosmic-ray phenomena. The main reason for creating such a project was that the cosmic-ray ensembles (CRE) are beyond the capabilities of existing detectors and observatories.  
Until now, cosmic ray studies, even in major observatories, have been limited to the recording and analysis of individual air showers therefore ensembles of cosmic-rays, which may spread over a significant fraction of the Earth  were neither recorded nor analyzed. 
In this paper the status and perspectives of the CREDO project are presented.
\end{abstract}
  
\maketitle

\section{Introduction}
\begin{figure}[h]
\centerline{%
\includegraphics[width=12.5cm]{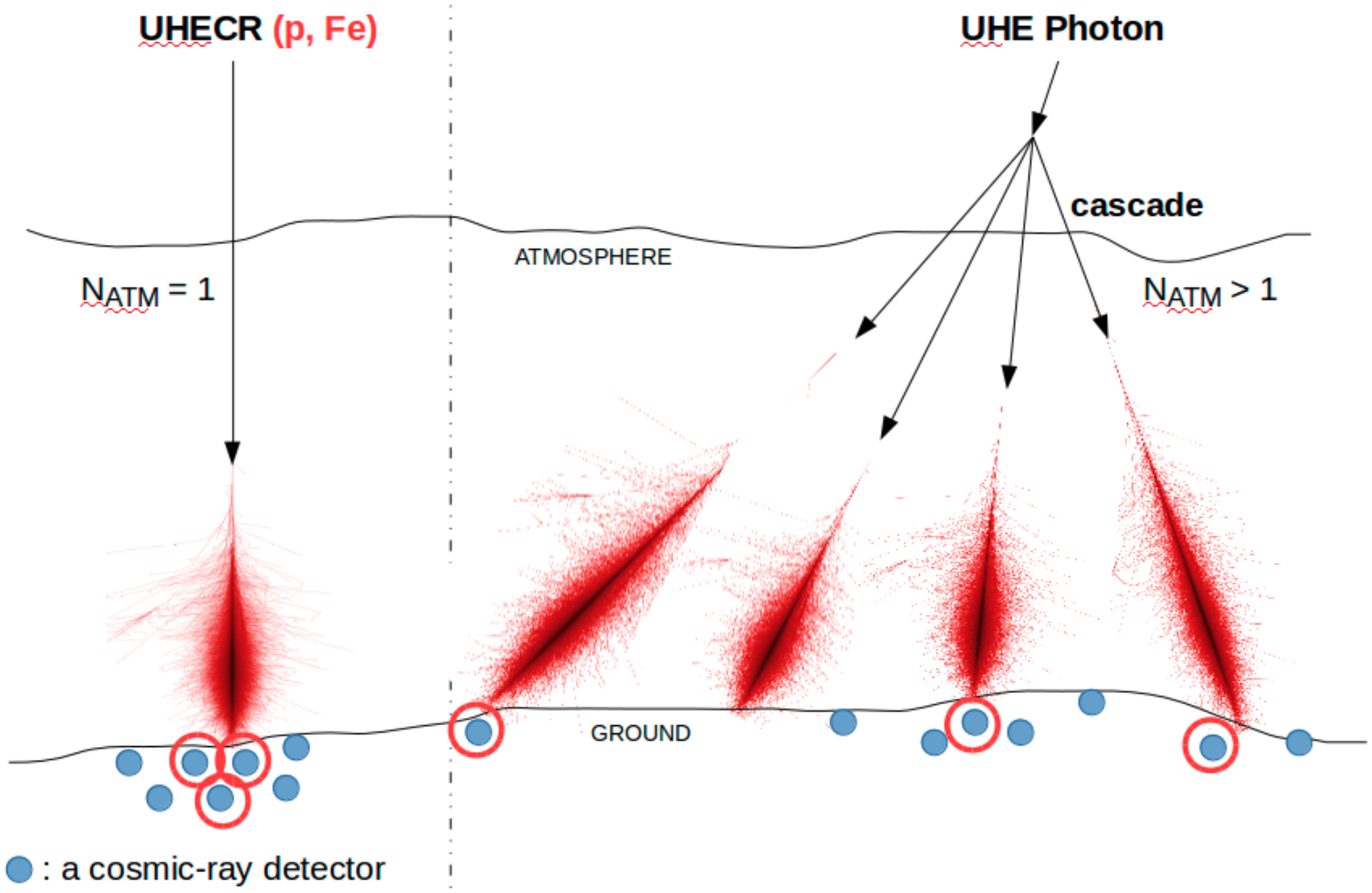}}
\caption{Left: Typical strategy: search for one extensive air shower (EAS) - cascade of secondary particles initiated by a single high-energy cosmic ray; Right: Cosmic-Ray Ensembles: a novelty in cosmic-ray
research and target for CREDO.}
\label{Fig:Strategy}
\end{figure}

The main goal of the CREDO project is to combine existing cosmic-ray detectors (large professional arrays,  educational instruments,  individual  detectors e.g.  smartphones, etc.)  into  one  worldwide  network, enabling global analysis of both individual cosmic showers and CRE presented on Fig. \ref{Fig:Strategy}.  
A very important aspect of this project is geographical  spread  of  the  detectors and access to big man power necessary to deal with vast amount of data to search for evidence for cosmic-ray ensembles.
This is due to the fact that the apple of the eye of this project and an integral part of the chosen  research method is citizen science enabling the use of work and the involvement of a huge number of people who are not necessarily scientists.

How a CRE can be produced? A good candidate is a shower induced by an ultra high energy photon interacting at least at some distance from the Earth e.g. close by the Sun. 
Such cascades, which are called preshowers \cite{PH2005}, are a consequence of an interaction of ultra high energy photons (UHE, with energies larger than $10^{17}$~eV) with solar or terrestrial magnetic fields. 
When the cascade arrives at the Earth, it comprises several thousand photons and
leptons with a peculiar spatial distribution to which CREDO will be sensitive \cite{ND2018}. 
Another possible mechanism of the production of CRE can be decay of long-lived
super-massive particles (M~$>10^{20}$~eV) and may lead to a significant fraction of UHE photons \cite{Chung} in cosmic ray flux and thus CRE.

CREDO proposes operation under a planetary network, what is necessary for signatures of CRE which may be spread over very large surface ($\approx 1000^2$~km).
The components of CRE might have energies
that practically span the whole cosmic-ray energy spectrum. 
Thus, all the cosmic ray detectors
working in this range, beginning from smartphones
and pocket scintillators, through numerous larger educational
detectors and arrays to the professional infrastructure that will receive cosmic rays as a signal or as a background could contribute to a common effort towards a hunt for CRE.

\begin{figure}[htb]
\centerline{%
\includegraphics[width=12.5cm]{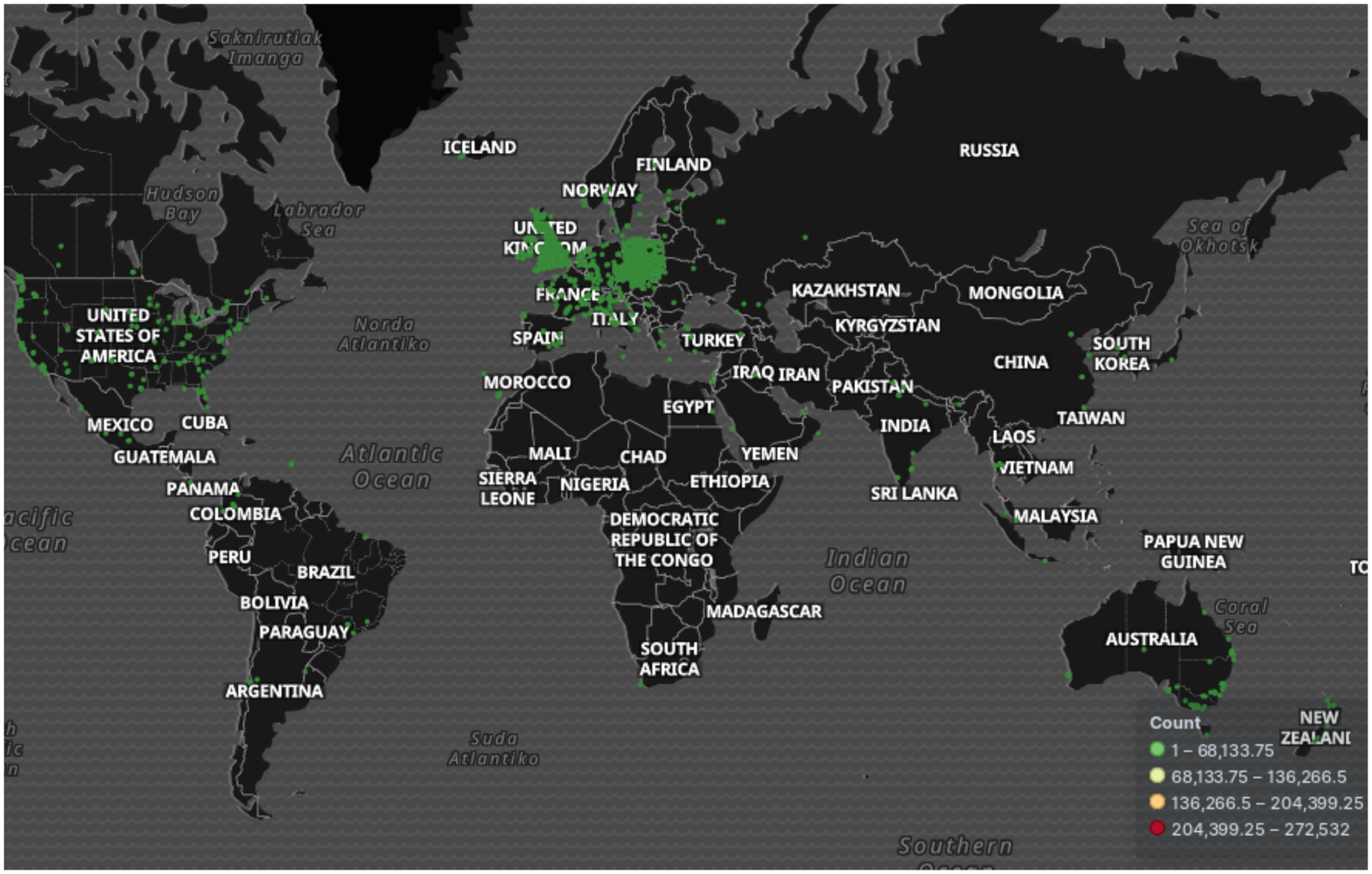}}
\caption{The map of CREDO user locations as of middle July 2019. More than 7500 users spread over the globe}
\label{Fig:Globe}
\end{figure}

In July 2019 number of institutions which are part of the CREDO collaboration was 23 located in 11 countries \cite{Collaborators2019}.
Important is number of active users of CREDO application (for smartphones), Fig. \ref{Fig:Globe} presents state of art in July 2019.

\section{Method}
Analysis of scenarios for the origin of cosmic particles that can be verified on Earth through an observation of an ensemble of at least two particles or photons with a globally spread and
coordinated network of detectors requires special and well-developed methods. 
The general CREDO strategy includes searches for groups of spatially correlated cosmic-ray photons that might arrive at the Earth at significantly different times,
with temporal dispersion of the order of minutes or more. 
Such phenomena have been reported in
the literature \cite{Obs1,Obs2}, but they have not been observed repeatedly until now. 
Of course spatial correlations of particles arriving
simultaneously at the Earth must be tested at the very beginning.

Cosmic-ray data are available everywhere (also inside buildings) they are free,
and they can be acquired with the minimum detection effort in an easy reach of the public - through
a mobile device equipped with an application that turns it into a particle detector. 
CREDO is going to use the already existing applications, the CREDO Detector \cite{CREDOdetector}, with the source code open under the MIT license that gives a freedom of development driven
even by a wide community, and unlimited flexibility to implement all the features essential for the
project. 
The other unique feature of the app is that it connects the user to the open server-side data
processing, analysis, and visualization system \cite{CREDOweb}, with open access to receive data.
It concerns also detectors other than smartphones, and opens access to the data in close to real time.

All the CREDO related codes developed so far are available on GitHub
repository \cite{CREDOgithub}.  
It enables an easy teaming up and continuation of the existing projects together with active participation and engagement of non-experts.

\vspace{-0cm}

\begin{figure}[h]
\centerline{
\includegraphics[width=18.5cm]{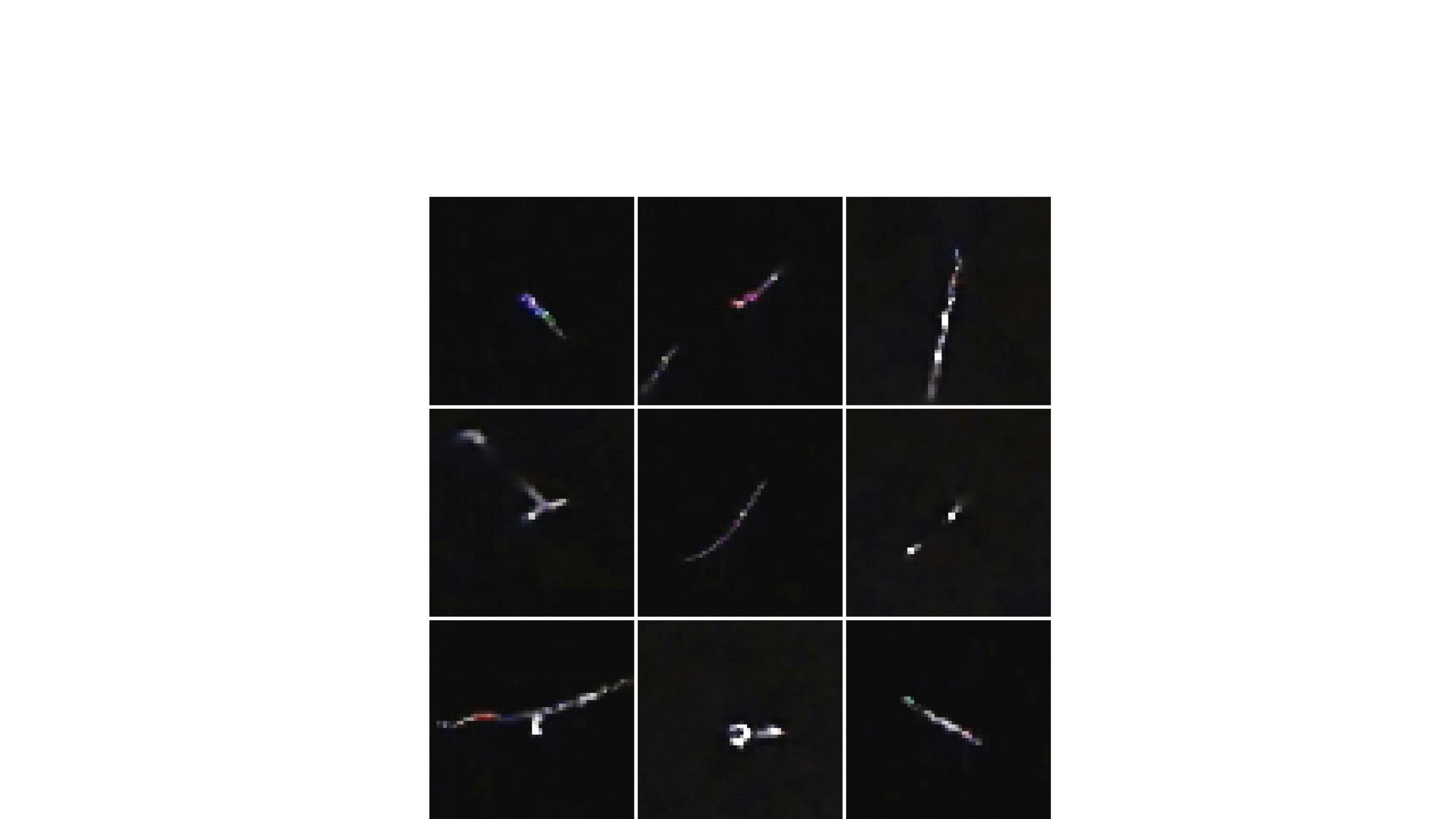}}
\caption{Tracks of particles detected by the CREDO Detector application}
\label{Fig:Tracks}
\end{figure}

Since a smartphone with the CREDO Detector is capable of detecting particle track candidates
(see Figure \ref{Fig:Tracks} for examples) including both the cosmic radiation (e.g. the penetrating particles like
muons) and the local radiation (e.g. X-rays) it opens a door to the public engagement model which
is based not only on the data classification, like for most of the citizen science projects, e.g. in
zooniverse.org, but also on the data acquisition.

Figure \ref{Fig:Globe} shows the map of the CREDO user's locations, from which one can see that
CREDO smartphones network already spreads over the planet. The number of registered users,
with at least one detection, in the middle of July 2019 was 7500 and about 2 918 000 images were
stored in the database. The observing time for all users equates to 958 years searching for particles,
which shows the large potential of such observations.

The unique scientific advantages that the smartphone cloud has in comparison to other detector
systems is the geographical spread and the public availability at no additional investment. 
These two features make the smartphone cloud a critically important and efficient component of the whole
CREDO system.

\section{First scientific results}

\subsection{Do CREDO smartphones really register single cosmic ray muons?}

Among the images in the CREDO database there are photographs of events where they are visible very long traces which are supposed to be images of tracks of cosmic ray muons that passed the matrix of the smartphone camera with large angles. 
The distribution of arriving zenith angles of single, incoherent muons is known for years. It is still a subject of measurements by the small, even portable muon telescopes often created and operated by students \cite{Wang1983, Franke2013, Sigh2015, Hutten2017}. 
The data used to obtain the results presented below consist of about 5$\%$ of registrations from the CREDO database collected in a long series on several different phones by two Collaboration members. In the database they are available as PNG files, cut from the whole camera frame to $60\times60$ pixels box around the lightest pixel in the whole frame. 

\begin{figure}[h!]
\centerline{
\includegraphics[width=11.cm]{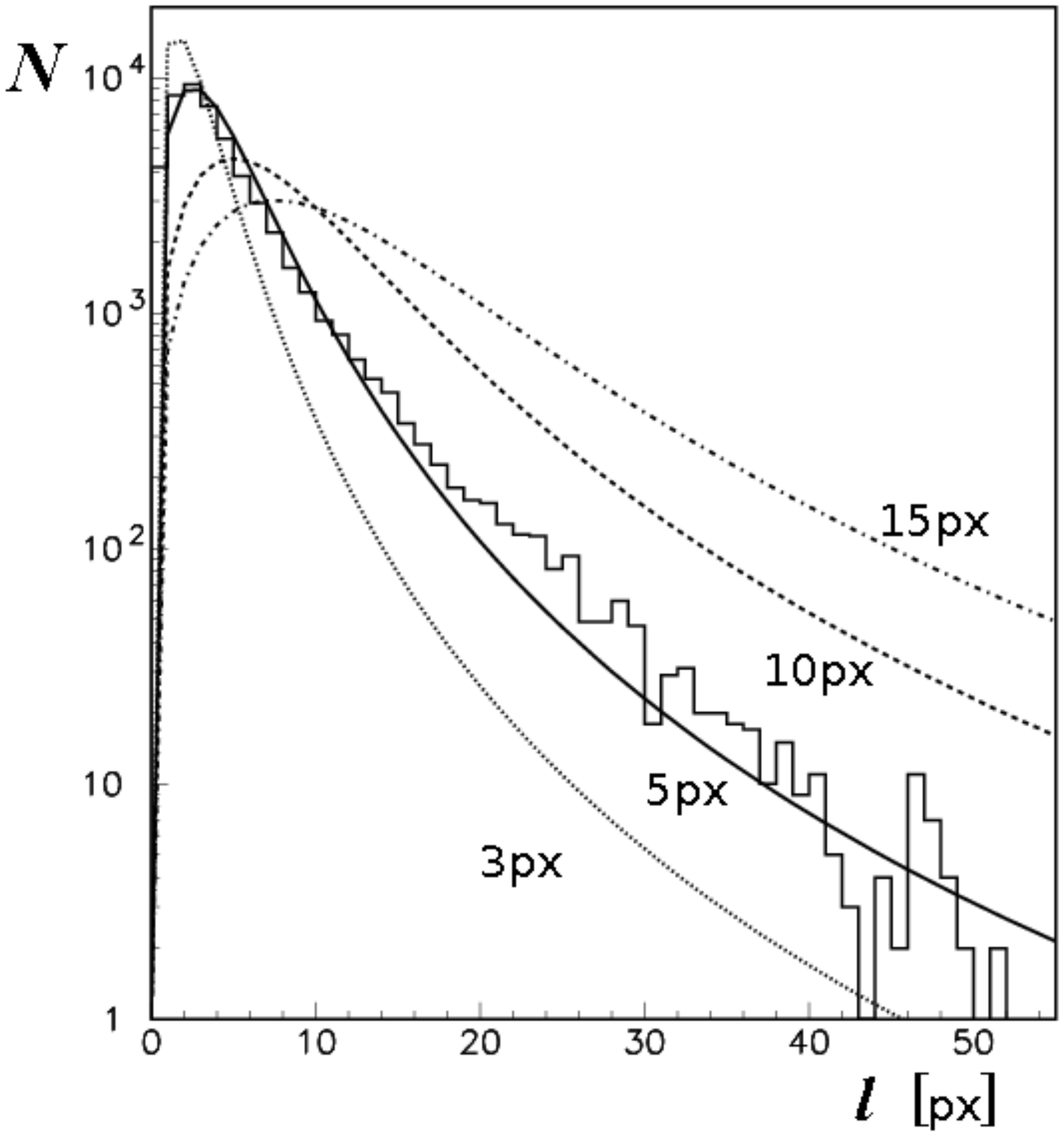}}
\caption{Measured track length distribution measured. The lines show comparison with the predictions for the various of the value of camera matrix height (h): 3 pixels -- dotted line, 5 pixels -- continuous line, 10 pixels -- dotted line and 15 pixels dotted line.}
\label{Fig:TracksLength}
\end{figure}

The crucial step in the analysis is to determine the main symmetry axis of the potential track. First  all pixels that exceeded the average noise (by ten times its average dispersion) are chosen and for them the main axis€™ of the track is determined. There are, in principle, many various possibilities to find it.  CREDO tested inertia ellipse, the Hough algorithm line, the smallest sum of squared distances weighted by squares of the brightness or by just the brightness. They all gave slightly different results, but the images of the tracks in the pictures are clear and regardless of how they are linearized, the 'large axis' of the track is almost always the same. Divergences sometimes appear in "multi-track" cases.
The distribution of the track length is shown as the histogram in Fig.~\ref{Fig:TracksLength}.

\vspace{0.5cm}

The track length $l$ relates to the zenith angles of incoming particles:
\begin{equation}
\Theta = ArcTan(l/h)
\end{equation}
where $h$ is the depth of the active layer of the matrix in the smartphone camera. The thickness of this layer is not exactly known and the cameras used for recording were different leaving the value of  a relatively 'free parameter' which effective value can by adjusted using the data.
The muon zenith angle distribution is known for 80 years and many different measurements confirm that it can be accurately described as
\begin{equation}
\frac{dN(\Theta)}{d\Theta} = cos^\gamma(\Theta)
\end{equation}
with the index of $\gamma \sim 2.0$.
Fig. \ref{Fig:TracksLength} shows that the values of  above a dozen pixel sizes or so are rather unacceptable (if we are measuring real cosmic ray muons), because they are not able to explain the observed large number of short traces. For the value of $h = 3$  one can see the opposite: the number of long traces seen is definitely too large. The value of the camera thickness of $\sim 5$ pixels quite well describes the measured distribution. 
One can conclude this analysis with the statement that the distribution of the zenith angle of particles responsible for the emergence of this tracks is in concordance with the expected distribution of the arrival angle of the single, incoherent cosmic ray muons.

\subsection{CREDO: EAS measurement}
However, the most interesting and attractive would be studying the phenomenon of Extensive Air Shower (EAS), a cascade of elementary particles, mostly photons and electrons, but also some muons or even high energy hadrons traveling almost at the speed of light from the upper atmosphere to the surface of the Earth. They arrive as a disk of millions of particles for one short instance. The source of very high energy particles that initiated such showers is, in general, not known as well as the mechanism of their acceleration from astrophysical sources. The mystery of (high energy) cosmic rays has stood for
almost a century and observing and studying these Extensive Air Showers can be exciting and stimulating for the young minds. 
The CREDO EAS array, called CREDO-Maze, was constructed and the prototype has registered its first showers. The concept of the CREDO-Maze array was developed based on the 20 years old Roland Maze Project \cite{Gawin2002, Feder2005}. The technology today has developed greatly and the local shower array idea of Linsley (1985) \cite{Linsley1985} can now be implemented much more easily and, critically, much more cheaply. Eventually CREDO wishes to present high-schools with sets of at least four "professional" cosmic ray detectors connected locally and forming the small school EAS array. These will use plastic scintillators instead of smartphone cameras, bespoke fast electronics instead of the smartphone application, and a more convenient connection to the wider CREDO database. But this is not all. CREDO established some physical arrangements that can be used by physics teachers in the standard physics education course showing properties of elementary particles, radiation attenuation, effect of interaction of particles with matter etc. There are many historical experiments (dating back to the 40s) that can be repeated in the classroom or during after hours activities. 
The final design of the small local EAS array is still in development, but to test the working principle CREDO have used small detector available on the market: the CosmicWatch Desktop Muon Detector. One set of Cosmic Watch contains two small (5cm x 5cm) scintillators monitored by silicon photomultipliers (SiPM) and slow electronics based on an Arduino microcontroller, that allows students to connect them to the computer and analyze simultaneous registration of signals from both detector laying one above the other, which flags muon passing through both scintillators.
CREDO have used two CosmicWatch sets, that is four individual scintillation detectors, bypassed the CosmicWatch electronics and used the raw SiPM signal. 
Four available individual detector signals into 6 pares of fast coincidence units have beed combined. The coincidence window width of 100 ns was used. This short time ensured that the signal rate from uncorrelated muon (and other noise sources) was of order of tens per minute to ensure that non-EAS signals will not disrupt the shower registrations. The logic of the array is such that if any coincidences appear in all 6 outputs, the data is immediately stored in a fast register and the trigger starts the slower electronics (based on an Arduino microcontroller). This last stage reads the register and transfers all 6 bits to the computer where they are stored together with the actual time stamp in a file for further analysis.

\subsection{Measurements}

Before using the array to search for an EAS, it was first tested to be sure that the registered events were real. 
The first test was to place four detectors in one tower arrangement. The CosmicWatch detectors are expected to give a signal for single muons. The threshold was set for all detectors€™ individual signal forming amplifier/comparators on the same level of about 30 mV chosen by comparing CosmicWatch original counting rate with the signal rate. 
The single muon passing at least two detectors can trigger the first and the second detectors, looking from the top of the tower and nothing more, leaving the tower from the side: (1, 2), can travers (1,2,3), (1,2,3,4), but also (2,3), (2,3,4) and (3,4). Other possibilities for the ideal geometry are forbidden.
The extremely low count for forbidden coincidence (with the "gap" between the fired detectors) configurations gives confidence that in the majority of cases CosmicWatch detector in the array respond correctly to the passage of the single relativistic charged particle.

\begin{figure}[h!]
\centerline{
\includegraphics[width=11.cm]{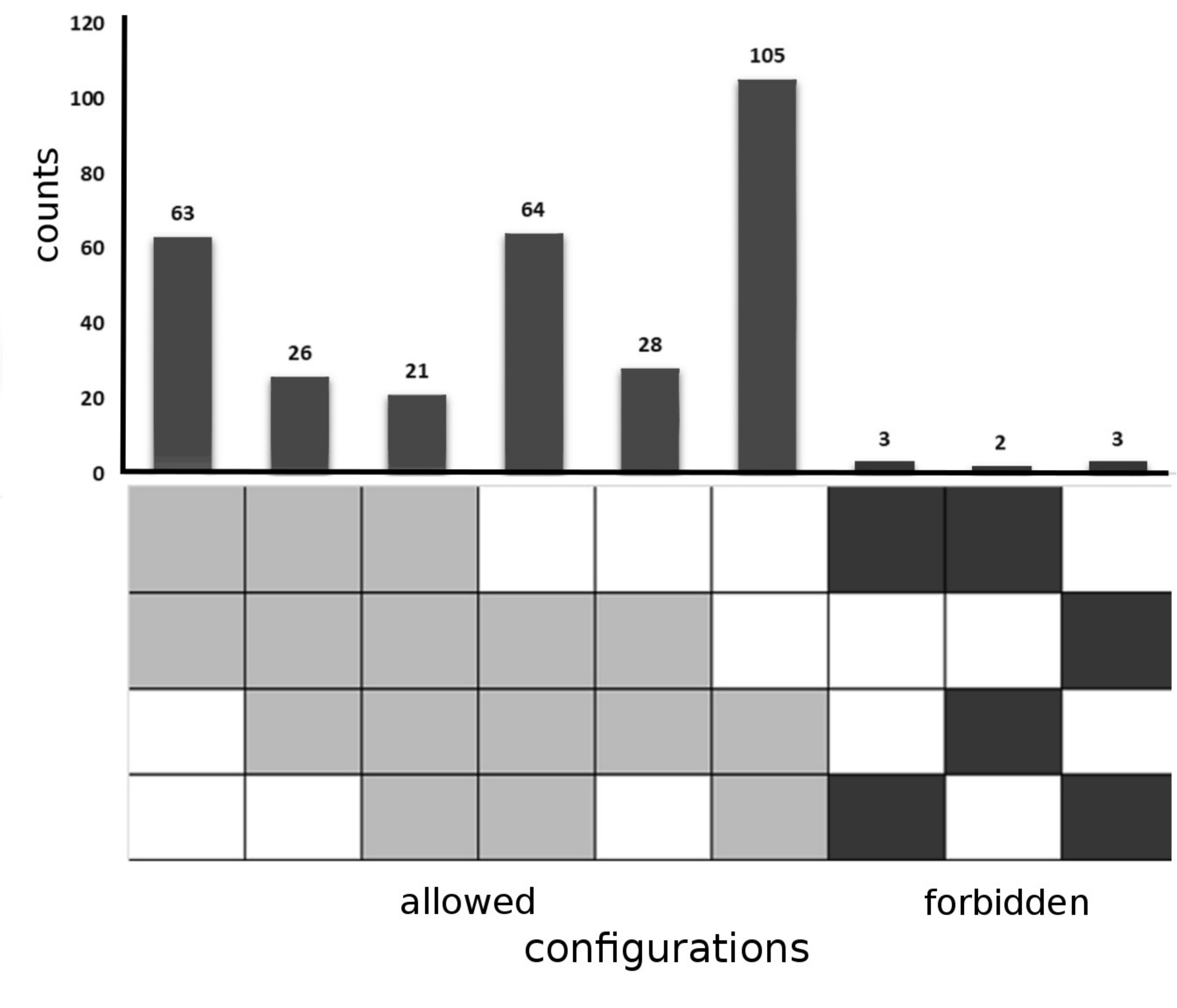}}
\caption{Rate of different coincidences recorded in the tower geometry}
\label{Fig:Counts}
\end{figure}

The asymmetry between first (1,2) and sixth (3,4) configurations, and (1,2,3) and (2,3,4) gives as a signal that the detector thresholds are not set perfectly. The number of counts shown in Fig. \ref{Fig:Counts} suggests that the detector number 1 seems to have the threshold set a little too high. However, this imperfection does not affect the general reasonable behavior of the whole array.
After the test, four detectors in the plain with irregular (not rectilinear) configuration of the step of order of 1 meter were arranged. Different geometries were tested without significant differences.  
The particles which could produce coincidence in the mini array have to come in one instant (within 100 ns) at the surface of about $1 m^2$. Considering the small effective area of CosmicWatch detectors, the particle density of particles in such event has to be rather high, or if it is actually low, the rate of such low-density events should be very high. Respective simulation calculations were performed. 
Millions of test densities according to one of the known examples of measured shower particle densities available on the market \cite{Broadbent1950} were generated. For each density was checked if each single detector was triggered or not. 

The single hits were, of course, lost in the background of single muons counts. In spite of the normalization, which is most problematic here, the ratios of three-fold to two-fold and four-fold to two-fold coincidences rate are observables which are  independent on the overall normalization (actual solid angle, shielding etc.) and depend mostly on the density spectrum index. 

The mini-array was active for just over a week, and during this time  94 two-fold coincidences, 2 three-fold and one all four hits event were registered.

The number of more than 2-fold coincidences is much too low, even if taking into account absolutely insignificant statistics. To perform an accurate measurement it was necessary to have detectors which respond in the same way to particle traversing the scintillator, which is not exactly this case (see Fig. \ref{Fig:Counts}).
Nevertheless, the comparison of the simulation result with made measurement gives confidence that the CREDO-Maze mini array registered real Extensive Air Showers, and one can say that it has been shown that the CREDO-Maze mini array using CosmicWatch detectors registered first Extensive Air Showers.

\section{Conclusions}
Pursuing the research strategy proposed in CREDO project will have a large impact on astroparticle
physics and possibly also on fundamental physics. If CRE are found, they could point back to
the interactions at energies close to the Grand Unified Theories (GUT) scale. This would give
an unprecedented chance to test experimentally for example dark matter models. If CRE are not
observed it would valuably constrain the current and future theories. Apart from addressing fundamental physics questions CREDO has a number of additional applications: integrating the scientific
community (variety of science goals, detection techniques, wide cosmic-ray energy ranges, etc.),
helping non-scientists to explore Nature on a fundamental but still understandable level. The high
social and educational potential of the project gives confidence in its contributing to a progress in
physics.

\vspace{1cm}

{\bf Acknowledgments}\\
The work was created as a result of the implementation of the Polish research project No. 2018/29/B/ST2/02576 financed from the funds of the National Science Center. This research has been supported in part by PLGrid Infrastructure. We
warmly thank the staffat ACC Cyfronet AGH-UST, for their always helpful
supercomputing support.

\end{document}